\documentclass{modified}

\begin{document}

\title{MONTE CARLO SIMULATION FOR SPHERES\\
WITH TWO LENGTH SCALES
}

\author{\footnotesize E.V.R.CHAN}

\address{ University of Washington\\
Seattle, Washington, 98195-2420, United States.\\
evr@u.washington.edu}

\maketitle

\begin{abstract}
Canonical ensemble Monte Carlo calculations have 
been carried out on spheres with two different
length scales.  Radial distribution functions at
various temperatures and densities were computed
and compared.  Preliminary results indicate that
there are differences but because these may be
subtle, more calculations including self-diffusion
coefficient,heat capacity,etc. would be necessary
in order to determine a phase diagram.
\end{abstract}

\begin{figure}[th]
\centerline{\psfig{file=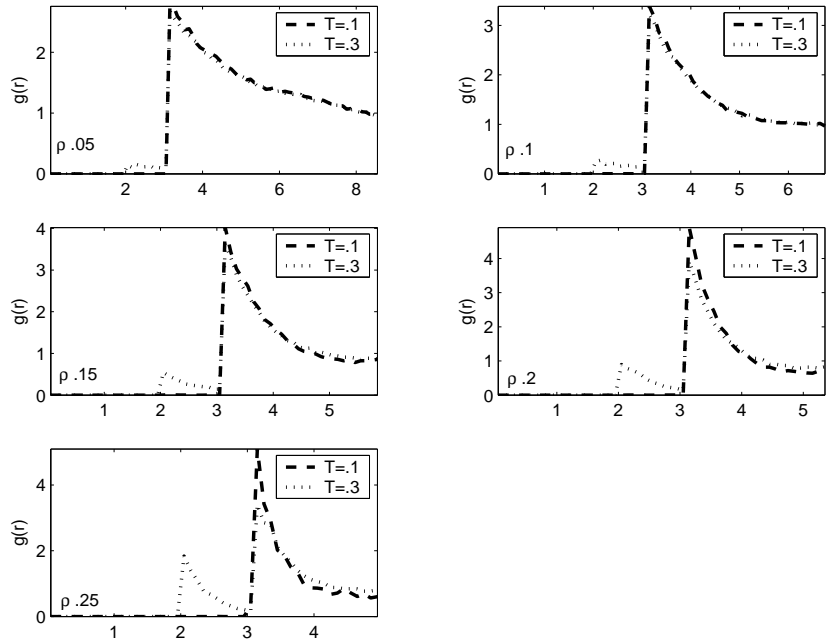,width=11cm}}
\vspace*{8pt}
\caption{ Radial distribution function g(r)
versus radius for densities .05,.1,.15,.2,.25
and temperatures .1 (dashed line) and .3 (dotted
line).  Density and temperature are in reduced units.  
At the lower densities the box the particles are 
enclosed by is bigger and so the radius extends
out further at the lower densities.}
\end{figure}

\begin{figure}[th]
\centerline{\psfig{file=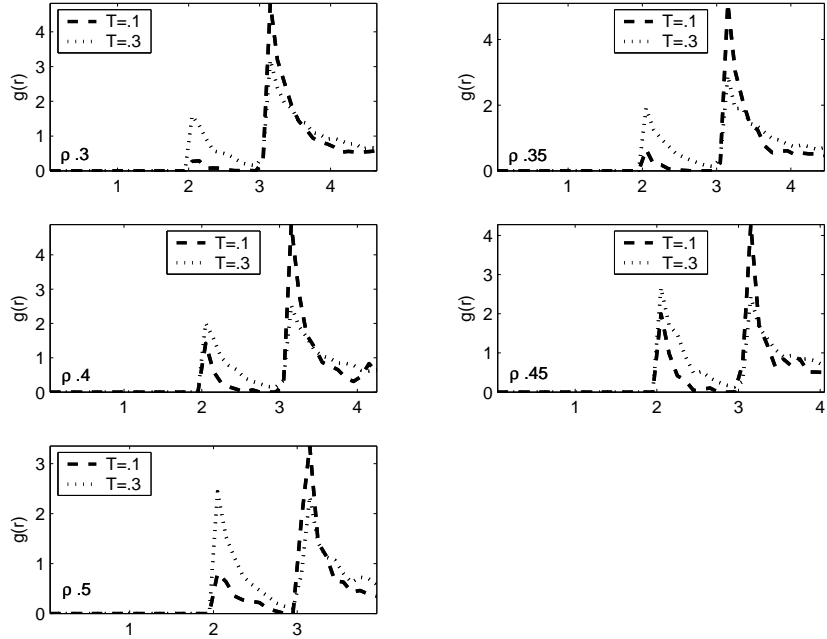,width=11cm}}
\vspace*{8pt}
\caption{ Radial distribution function g(r) versus 
radius for densities .3,.35,.4,.45,.5 and temperatures
.1 (dashed line) and .3 (dotted line).  Density and
temperature are in reduced units.  At the lower 
densities the box enclosing the particles is bigger
and so the radius extends out further at the lower 
densities.}
\end{figure}

\section{Introduction}

Repulsive step potentials, or collapsing sphere$s^1$,
have been used to describe liquid-liquid phase 
transition$s^2$ and pattern formation in self-
organizatio$n^3$,  The TIP5P model of wate$r^4$ has
a steep repulsion on the oxygen (part of the Lennard-
Jones potential) and Coulombic potentials centered on 
the hydrogens plus two other points, all tetrahedrally
located from the oxygen, so that near the oxygen there 
is some kind of a step potential.

\section{Experiment}

Simulations using a computer program written to calculate
the radial distribution function of hard spheres with a
repulsive step were done.  The potential is characterized
by sigma=2=hard sphere diameter and sigma1=sigma*1.55=3.1=
diameter of second length.  The step lies between sigma 
and sigma1 and has height Ep.  For distances greater than
sigma1 the potential is zero and less than sigma, is 
infinite. The temperature used (redT in the program) is in
reduced units kT/Ep, where  k is the Boltzmann constant.

The reduced density is given by the number of particles
divided by the volume of the box (N/V) and this is multiplied
by the hard sphere diameter cubed. The number of particles 
used is reporte$d^5$ to be as low as 32.  Recen$t^6$
research indicates that size effects using 32 particles
are small.  Thirty two particles were used in a canonical
(NVT constant) ensemble for this preliminary study.  
The number of equilibration cycles equalled the number of
measurement cycles of 100,000 MC cycles.  One MC cycle 
or one Monte Carlo step per particle (MCS),
is the equivalent of the "time" it takes for N particles
on the average to have had a chance to change their
coordinates.  
As density increased in Figures 1 and 2, the peaks
were getting sharper; this is characteristic of greater 
order among the particles.  Structural 
changes in the arrangements of the spheres are 
expecte$d^7$ to be exhibited by changes in the radial
distribution function.

\section{Conclusions}

Preliminary results indicate that there are changes
in g(r) brought about by structural changes, however
some of these can be subtle.  It is best to get more
thermodynamic data including  mean square displacement 
and specific heat in order to determine the phase 
diagram.



\begin{thebibliography}{0}
\bibitem{1}
V.N.Ryzhov and S.M.Stishov,
{\it Phys. Rev. E} {\bf 67}, 010201(R) (2003).

\bibitem{2}
P.J.Camp, {\it Phys. Rev. E} {\bf 68 }, 061506 (2003). 
\bibitem{3}
G.Malescio and G.Pellicane,
{\it Nature Materials} {\bf 2}, 97 (2003).

\bibitem{4}
M.W.Mahoney and W.L.Jorgensen,
{\it J. Chem. Phys.} {\bf 112} 8910 (2000).

\bibitem{5}
J.A.Barker and D.Henderson in
{\it Annual Review of Physical Chemistry},
{\bf 23}, 439 (1972).

\bibitem{6}
E.Schwegler, J.C.Grossman, F.Gygi and G.Galli, "Towards an
Assessment of the Accuracy of Density Functional Theory
for First Principles Simulations of Water II, arXiv:
cond-mat/04055611.

\bibitem{7}
T.M.Truskett, S.Torquato, S.Sastry, P.G.Debenedetti 
and F.Stillinger, {\it Phys. Rev. E } {\bf 58 },
3083 (1998).



\end{thebibliography}
\end{document}